# Robustness of indispensable nodes in controlling protein-protein interaction network

Xizhe Zhang[1], Huaizhen Wang[1], Yunyi Yang[2]

[1](School of Computer Science and Engineering, Northeastern University, Shenyang110819, China)

[2](IT Center, National Audit Office, Beijing 100073, China)

**Abstract:** Recently, the structural controllability theory has been introduced to analyze the Protein-Protein Interaction (PPI) network. The indispensable nodes, which their removal increase the number of driver nodes to control the network, are found essential in PPI network. However, the PPI network is far from complete and there may exist many false-positive or false-negative interactions, which promotes us to question: are these indispensable nodes robust to structural change? Here we systematically investigate the robustness of indispensable nodes of PPI network by removing and adding possible interactions. We found that the indispensable nodes are sensitive to the structural change and very few edges can change the type of many indispensable nodes. The finding may promote our understanding to the control principle of PPI network.

## Introduction

The protein-protein interaction (PPI) network [1-3] are essential to analyze many biological processes [4, 5] and protein functions [6, 7]. The PPI network can be obtained by high throughput yeast two-hybrid screens (Y2H) [2]. The analysis of PPI network provide many insight into protein function and identifying important protein nodes [8].

Recently, the structural controllability theory [9] has been introduced to analyze PPI network. One approach [10-12] assumed that the protein nodes can control all its edges independently, and the minimum set of nodes used to control the PPI network can be obtained by the minimal dominate set of a network. Another approach [13-15] assumed that the node can control only one of its outgoing links. The minimum set of driver nodes, which are used to full control the network, can be obtained by the maximum matching, and the unmatched nodes are the driver nodes. As the nodes of real networks can be disabled or removed from the network, Liu et.al [16] classified the nodes based on their influence to the maximum matching. If a node's removal decrease the size of maximum matching, it is called indispensable node; if the size of maximum matching is increased, it is called dispensable node; and if the size of maximum matching are unchanged, it is called neutral node. The above classification of nodes represent the roles of nodes participate in the control process.

Based on above node classification, Vinayagam et.al [17] analyzed the node type of PPI network [18]. They found that the indispensable nodes were the primary targets of disease genes and drugs, and the indispensable nodes can be used to identify new target diseases such as cancer genes. However, the PPI network is far from complete [19] and there may have many undiscovered interactions. Therefore, it is essential to evaluate the robustness of indispensable nodes. Furthermore, the PPI networks also have many false-positive interactions which rooted in the technical limitation. Although there are some results [17] about robustness of indispensable nodes based on the edge removal, we still feel it is insufficient and worth to further investigation.

Here we analyzed the robustness of indispensable nodes of PPI network by removing or adding edges. We counted the number of unchanged indispensable nodes after adding or removing edges from PPI network, and found that the indispensable nodes are sensitive to structural change. Surprisingly, we found that for some well-connected sub-graphs, in which the indispensable nodes are extremely sensible to the structural changes. Therefore, the indispensable nodes may be not a good indicator to identify important nodes of PPI network.

**Result**

First we introduced some basic concepts about structural controllability of complex networks. A network is said to be controllable if its states can be driven from any initial state to a desired state. The nodes which connected to the external control signals are called driver nodes. For any maximum matching (an edges set which do not share nodes) of a network, it is proved [20] that the set of unmatched nodes are the minimum set of driver nodes. If a node's removal increase the size of *MDS*, we say it is an indispensable node. In another word, removing an indispensable node will decrease the size of a maximum matching. Figure.1 show an example network and the types of nodes. Because the indispensable nodes are determined by the maximum matching of a network, the change of edges may affect the type of indispensable nodes. For example, in Figure.1, node 4 is changed from indispensable to neutral after edge (2,7) added.

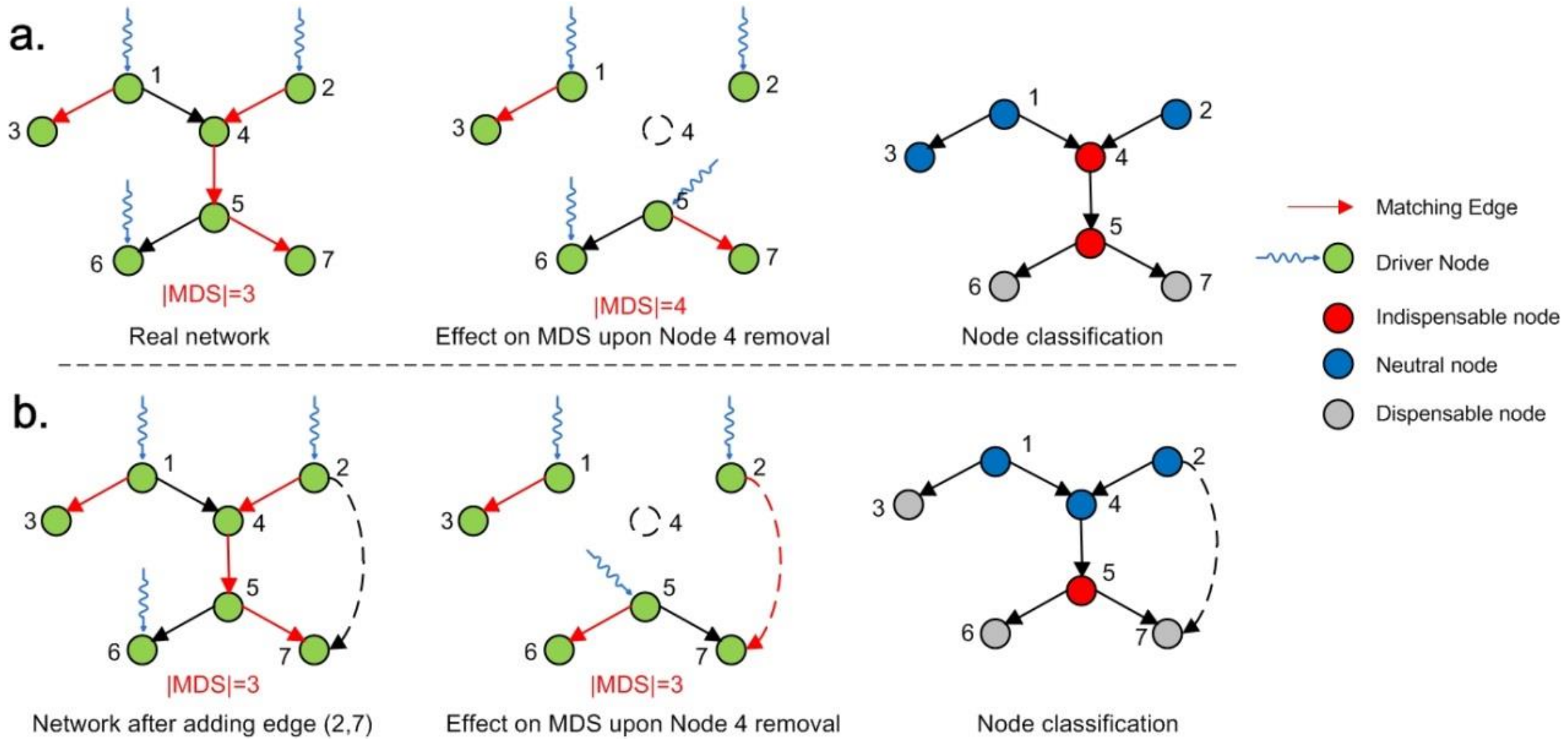

**Figure.1** | Indispensable nodes of a network may be sensitive to the structure changes. (a). a sample network with three driver nodes. Node 4 is an indispensable node because its removal increase the number of driver nodes, same as node 5; (b) a new network by adding edge (2,7) to the network shown in (a). Node 4 is no long indispensable because the number of driver nodes do not change after removing node 4.

To investigate the robustness of indispensable nodes, we used the directed human PPI network presented in [18], which has 6,339 proteins (nodes) and 34,813 interactions (edges). To evaluate the robustness of the indispensable nodes, we constructed a series of networks by removing or adding edges to the original PPI network. We tried to find the answer of the following questions: if we add or remove some edges of PPI network, how many indispensable nodes of original PPI network are still indispensable in the new networks? Let the set of indispensable nodes of original PPI network be $I_{ori}$, and the set of indispensable nodes of the new network be $I_{new}$. Therefore, the robustness of indispensable nodes can be evaluated by $R = (I_{ori} \cap I_{new})/I_{ori}$, that is the fraction of unchanged indispensable nodes after structural change.

### Removing edges

First we analyzed the robustness of indispensable nodes by removing edges from original PPI network. For the PPI network present in [18], each edge has a confidence score (CF) which represent the possibility of the existence of the edge. We removed the edges based on their confidence scores and found the fraction of unchanged indispensable nodes are significantly decreased with the edge cutoff score (Figure.2A). We found that about 58.9% indispensable nodes (784 of 1330) of original PPI network are no longer indispensable in the new network after removing all edges which *CF*<0.9, which indicate that indispensable nodes may not robust to removing edges.

Furthermore, we found that if the edges are removed based on some specific strategies, the indispensable nodes will be changed by very few edges. We already known that the indispensable nodes are not the driver nodes [17], and always be matched. Therefore, if we changed the matching status of an indispensable node, that is, let the indispensable node to be unmatched by any maximum matching, we will change the type of indispensable nodes. Based on above idea, we design an edge removing strategy to efficient change the type of indispensable nodes. The results (Figure.2B) show that the 39.2% indispensable nodes (522 of 1330) are changed by 7.7% removed edges (2700 of 34813). Note that the confidence score of all removed edges are still under 0.9. Therefore, we can change many indispensable nodes with very few removed edges.

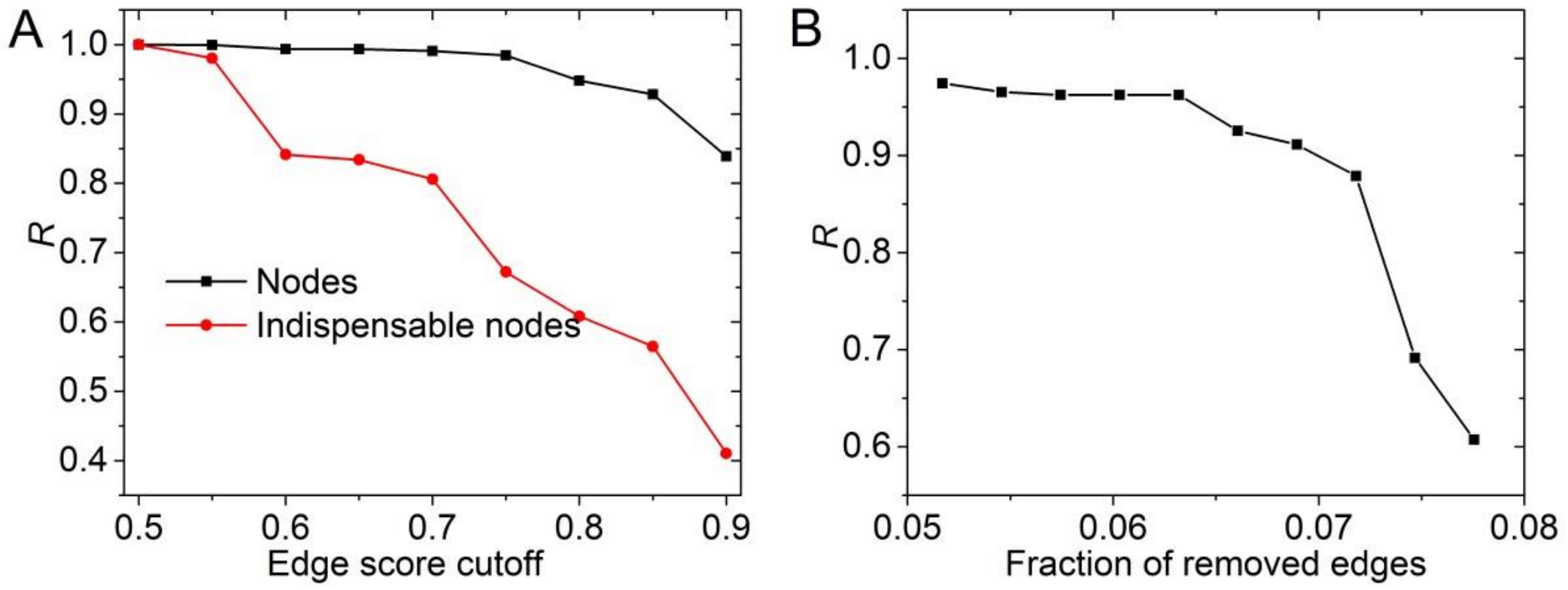

**Figure.2** | the fraction of unchanged indispensable nodes after removing edges. A. the fraction of unchanged indispensable nodes *R* is decreased with the cutoff score of edges. The edges are removed based on their confidence scores from 0.5 to 0.9; B. the fraction of unchanged indispensable nodes *R* versus fraction of removed edges. The edges which removed are selected based on their participation in maximum matching, that is, we removed the edges which connected the indispensable nodes with high priority.

### Adding edges

Next we analyzed the robustness of indispensable nodes by adding edges between the proteins. Surprisingly, we found that the indispensable nodes of some sub-graphs can be altered by only one added edge. Figure.3 showed an example sub-graph, in which contains 16 indispensable nodes. If we added an edge pointing from node MCM4 to any driver nodes, these indispensable nodes will be changed to neutral nodes.

The reason of these sub-graphs can be easily altered rooted in that the nodes are all matched.

Therefore, if we linked a unmatched node (driver node) to the sub-graph, the nodes of the sub-graph will turn into possible driver nodes , which is proved by recent work[21].The edges can connect any driver nodes to the neighbor of indispensable nodes.

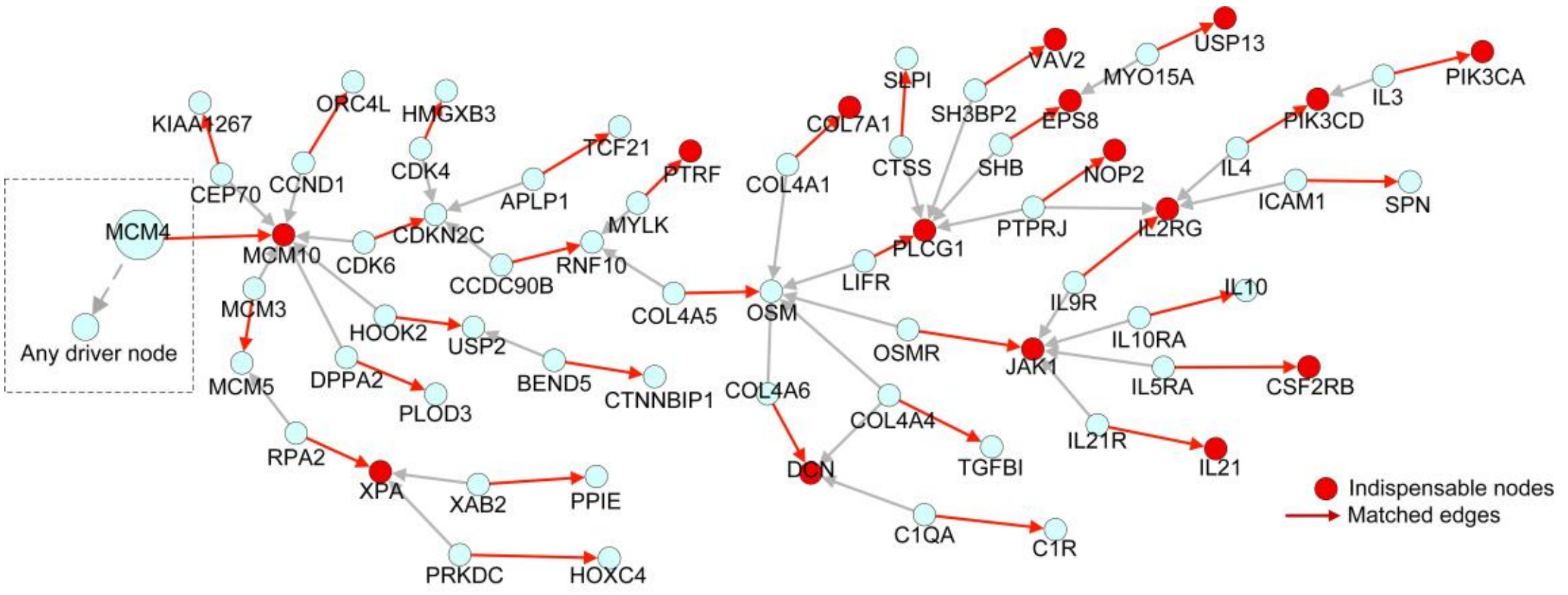

**Figure.3** | an example of sub-graph in the PPI network. The sub-graph contains 16 indispensable nodes, they can be changed by only one added edge, which is the edge pointing from node MCM4 to any driver nodes of the PPI network.

Furthermore, for the other parts of the PPI network which do not have perfect matching, we added edges to the network and made the unmatched node be matched, and then the sub-graph can be easily altered. Based on above strategy, we found that the fraction of three type of nodes are significantly changed with the number of added edges (Figure.4A), and very few added edges can change lots of indispensable nodes. For example, 89.2% indispensable nodes (1187 of 1330) of original network will be changed after 5.7% (2000 of 34813) edges added (Figure.4B). Surprisingly, the number of indispensable nodes is plunged when 4.8% edges added, indicating that there exist a giant sub-graph which can be easily changed after edges added.

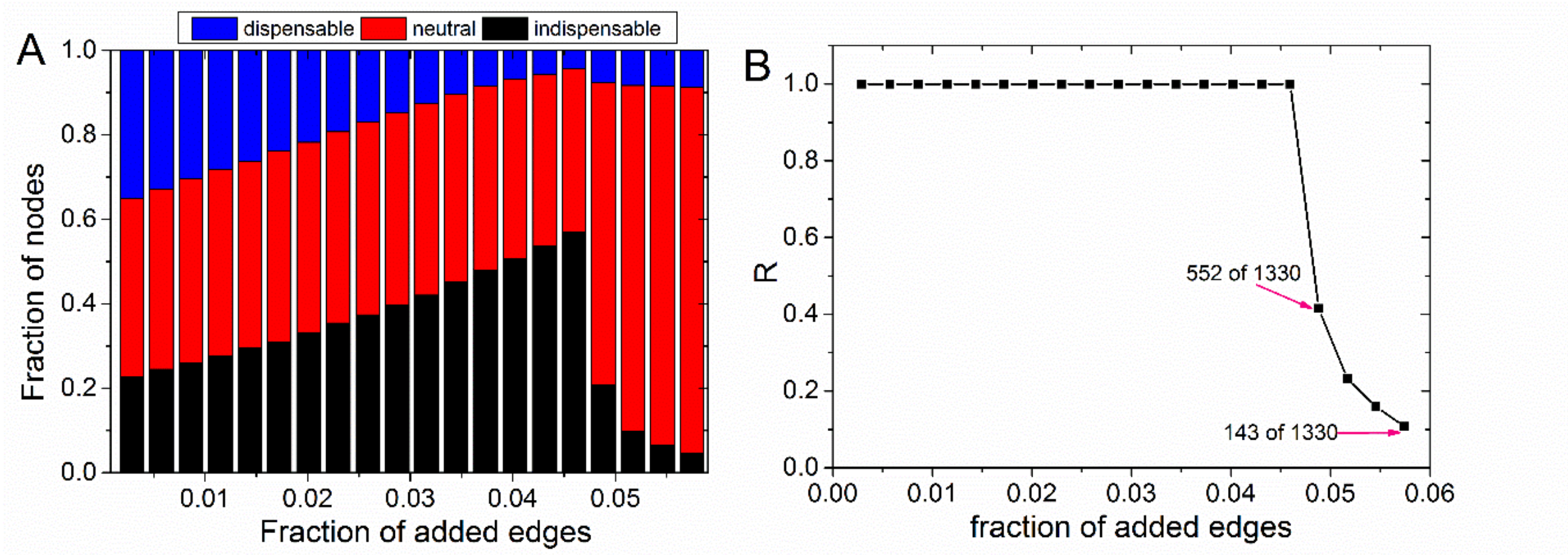

Figure.4 | the fraction of nodes versus the fraction of added edges. A. the fraction of all three type of nodes are changed with the number of added edges in the new network; B. the fraction of unchanged indispensable nodes is plunged with the fraction of added edges.

A major concern about the adding edge strategy is that the added edges may not exists in PPI network. However, the above adding strategy are not unique. In fact, for the sub-graph shown in Figure.3, any edge which connect any driver nodes (2283 nodes) and the neighbors of indispensable nodes will change the indispensable node and its descendant nodes. Therefore, there are many potential edges can change the sub-graph. Unless all these edges are not exist, the sub-graph will always be changed. Therefore, the indispensable nodes can be changed by few added edges with high probability.

## Discussion

In conclusion, the indispensable nodes of PPI network are sensitive to structural perturbation. The indispensable nodes are not the driver nodes, therefore, it is also a part of redundant nodes (nodes never act as driver nodes). Existing works [22] found that the fraction of reductant nodes exhibit bifurcation phenomenon in dense networks. Minor structural change may cause change of type of most of nodes in the network.

Since the PPI network are far from complete, it is reasonable to analyze the nodes are still have same properties after adding edges. The existing indicators used to identify the important nodes, such as degree and other centrality, are robustness to the structure change. That means that few added cannot significantly change the value of these indicator. For example, for a node with high degree, its degree will not significant change when few new edges are added.

However, the indispensable nodes are very sensible to structure change, very few added or removed edges may change the type of existing indispensable nodes. Therefore, the indispensable nodes may not a good indicator to identify the importance of node of the PPI network. The reason of why these indispensable nodes are enriched with disease genes and drug target reported in [17] may lie on some in-depth principles and worth to further investigation.

## Acknowledgments

Supported by the Fundamental Research Funds for the Central Universities of China under grand number N140404011 and by the Natural Science Foundation of China under grant number 60903009, 71272216, 91546110.

## Author Contributions

X.-Z.Z. designed research and wrote the paper. H.-Z.W. performed the experiments and analyzed the data. Y.-Y.Y. analyzed the data. All authors reviewed the manuscript.

## Additional Information

The authors declare no competing financial interests. Correspondence and requests for materials should be addressed to X.-Z.Z. (Email: zhangxizhe@ise.neu.edu.cn) or Y.-Y.Y. (Email: peteryoung@audit.gov.cn)